\documentstyle[12pt,epsf,epsfig]{article}
\def\btt#1{\documentstyle[eqsecnum,aps]{revtex}
{\tt$\backslash$#1}}

\begin{document}
\author{T. Tsang, V. Castillo, R. Larsen, D. M. Lazarus, D. Nikas,
C. Ozben, \\ Y. K. Semertzidis, and T. Srinivasan-Rao \\ 
Brookhaven National Laboratory, Upton, NY 11973 \\
L. Kowalski \\
Montclair State University, Upper Montclair, NJ 07043}
\title{Electro-optical Measurements of Ultrashort 45 MeV Electron Beam
Bunches}
\date{\today}
\maketitle
\begin{abstract}
We have measured the temporal duration of 45 MeV picosecond electron
beam bunches using a noninvasive electro-optical (EO) technique. The
amplitude of the EO modulation was found to increase linearly with
electron beam charge and decrease inversely with distance from the
electron beam. The risetime of the  temporal signal was limited by our
detection system to ${\rm\sim70\ ps}$. The EO signal due to ionization
caused by the electrons traversing the EO crystal was also
observed. It has a distinctively long decay time constant and signal
polarity opposite to that due to the field induced by the electron
beam. The electro-optical technique may be ideal for the measurement
of bunch length of femtosecond, relativistic, high energy, charged,
particle beams. 
\end{abstract}
\noindent
pacs: 07.77.Ka, 33.55.Fi, 78.20.Jq
\section{Introduction}

With the advance of electron and particle accelerators, the particle
bunch duration has dropped to the femtosecond time scale. Various
techniques have been proposed to measure such ultrashort bunch
lengths. One of the techniques relies on non-coherent transition
radiation where visible photons are collected and measured with a
streak camera.\cite{Qui} Although such a technique can yield bunch
length information, it is an invasive technique and the resolution is
sensitive to the photon collection system . Recently, the development
of the electro-optical probe based on the linear Pockels effect has
revolutionized the noninvasive measurements of small electronic signal
propagation on integrated circuits,\cite{Sheridan,Dykaar} dc and ac
high voltages,\cite{Santos,Rose,Murooka} lightning
detectors,\cite{Koshak} terahertz electromagnetic field
imaging,\cite{Jiang}and electron beam measurements of long
pulse\cite{Brubaker} and short pulse
duration.\cite{Geitz,Oepts,Semertzidis,Knippels,Fitch} 
EO sensors that use fibers for input and output coupling provide
excellent electromagnetic isolation and large frequency response,
limited essentially by the fibers and the velocity mismatch of the
electrical and optical waves to picosecond or sub-picosecond time
resolution. In this work, we show that the fast component of the EO
modulation is due only to the transient electric field induced by the
passage of an ultrashort relativistic electron bunch. No cavity
mode~\cite{Fitch} was observed. We examine the dependence of the
EO modulation with charge and its position. Finally, we present a
detection-limited temporal shape of EO signal and then draw our
conclusions regarding the electron bunch length. 

The optical probe is based on the principle of the linear
electro-optical effect - Pockels effect. When an electric field is
applied to a birefringent crystal, the refractive index ellipsoid is
modulated and an optical phase shift is introduced. To probe the phase
shift, an optical beam polarized at ${\rm 45^o}$ to the z-axis of the
EO crystal is propagated along the y-axis of the crystal. This phase
retardation is converted to an intensity modulation by a $\lambda
\over {\rm4}$ waveplate  followed by an analyzer (crossed
polarizer). The intensity of light $I(t)$ transmitted through the
analyzer can be described by\cite{Yariv}
  
\begin{equation}
I(t) = I_{o} [ \eta\ +\ {\rm sin}^2 ( \Gamma_o\ +\ \Gamma_b\ +\ \Gamma
(t))] \label{EQN(I(t))},  
\end{equation}

\noindent
where $I_o$ is the input light intensity, $\eta$ contains the
scattering contribution of the EO crystal and the imperfection of the
polarizer and other optics which is typically much less than 1,
$\Gamma_o$ contains the residual birefringence of the crystal,
$\Gamma_b$ is the optical bias of the system which is set at $\pi
\over 4$, and $\Gamma (t)$ is the optical phase induced by  the
electric field imparted on the crystal. When ${\rm \Gamma_o\ +
\Gamma_b \simeq {\pi \over 4}}$, Eq.(\ref{EQN(I(t))}) can be written
as 
\begin{equation}
{ {I (t)}\over I_o} \simeq (\eta + {1 \over 2}) + [{1 \over 2} {\rm
sin} (2 \Gamma (t))]  \label{I(t)/Io}  
\end{equation}
where the first term is the unmodulated dc light level which is
approximately equal to half of the input light intensity, and the
second term is the EO modulation. For a weak modulation, i.e. $2
\Gamma (t) \ll 1$, the EO component can be written as  
\begin{equation}
[{{I (t)}\over I_o}]_{\rm EO}\ \simeq\ \Gamma (t)\ . \label{EO}
\end{equation}
The normalized light output is approximately linear in the
time-dependent optical phase. 

The transient optical phase shift is linearly proportional to the
time-dependent field $E_z (t)$ traversing the optical axis of the EO
crystal and can be expressed as, 
\begin{equation}
\Gamma (t)\ =\ {1 \over 2}({n_e}^3 r_{33} - {n_o}^3 r_{13}){{2\pi L
E_z (t)} \over \lambda},  \label{Gamma(t)} 
\end{equation}
with $L$ the effective length of the crystal, $n_e$ and $n_o$ the
extraordinary and ordinary indices of refraction, $r_{33}$ and
$r_{13}$ the electro-optical coefficients, and $E(t)$ the transient
electric field in vacuum directed, along the optic axis (z-axis),
induced by the passage of the relativistic electron beam. This
relationship holds when the duration of the electric field is greater
than or equal to the time needed by the laser light to traverse the
entire length $L$ of the EO crystal.  

The electric field induced by a nonrelativistic electron beam is
radially symmetric. However a 45 MeV relativistic beam produces an
anisotropically directed radial field orthogonal to the electron beam
direction and along the z-axis of the EO crystal.\cite{Jackson} The
traverse field strength $E_z$ is given by  
\begin{equation}
E_z (t) = {1 \over {4 \pi \epsilon_o}}{{\gamma\ N_e\ q\ T(t)} \over
{\epsilon\ r^2}},\label{Ez} 
\end{equation}
with $\gamma$ the relativistic Lorentz factor, $N_e$ the number of
electrons in the beam, $q$ the electron charge, $T(t)$ the temporal
charge distribution, $\epsilon_o$ the permittivity of free space,
$\epsilon$ the dielectric constant of the EO crystal in the z-axis
direction, and $r$ the radial distance of the electron beam from the
axis of the optical beam. This electron beam field is present for a
time~\cite{Jackson}  
\begin{equation}
\Delta t = {r \over {\gamma \upsilon}},\label{Delta-t}
\end{equation}
with $\upsilon$ the electron beam velocity.  In this experiment
$\Delta t$ is approximately 100 fs which is much shorter than the
electron bunch length of $\sim {\rm 10\ ps}$. Thus, for an
uncompressed electron beam, ignoring any nonlinear beam dynamics, the
electron bunch length measurement is not distorted.  Also, in writing
Eq.(\ref{Delta-t}) one approximates the longitudinal size of the
electron beam to be negligible compared to the laser beam width. When
the  longitudinal size of the electron beam is larger than the laser
beam width, the electron charge that influences the optical phase of
the laser field is only (to first order) the fraction of charge over
the laser beam width. The actual strength of the electron beam field
is thus lowered due to this geometrical factor.  

The effective length $L$ of the crystal is the distance light travels
inside the crystal during the time $\Delta t$. When the electron
velocity $\upsilon$ approaches $c$, $L$ is given by 
\begin{equation}
L = \Delta t \times {c \over n} \simeq {r \over \gamma n},\label{L}
\end{equation}
with $n$ the index of refraction of the crystal at the laser
wavelength. Substituting Eq.(\ref{L}), Eq.(\ref{Ez}), and
Eq.(\ref{Gamma(t)}) to Eq.(\ref{EO}) gives 
\begin{equation} 
[{{I (t)}\over I_o}]_{\rm EO}\ \simeq\ ({n_e}^3 r_{33} - {n_o}^3
r_{13}) {{N_e\ q\ T(t)} \over {4\ \lambda\ n\ \epsilon_o\ \epsilon\
r}}.  \label{final-EO} 
\end{equation}
The optical phase is modulated only during the time the electron beam
field is present. However, the duration of the EO signal depends on
both the electron temporal charge distribution $T(t)$ and the length
of the crystal since all the light will be influenced at the same
time. Notice that Eq.(\ref{final-EO}) has no $\gamma$ dependence, it
depends linearly on electron charge $N_e q$, and it has a $1 \over r $
dependence and not $1\over r^2$. Furthermore, the $ 1 \over \epsilon $
dependence favors EO crystals with a small dielectric constant.   

\section{Experimental arrangement}
 
A vacuum compatible EO modulator setup was constructed using discrete
optical components mounted on an aluminum bar anchored to a standard
${\rm 2 {3 \over 4}}$ inch O.D. vacuum flange, see Fig. 1(a). The
complete setup was designed to fit into a conventional 1.37 inch
I.D. 6-way cross, and a 45 MeV electron beam passes through the center
of this vacuum beam pipe. The light source was a fiber-coupled,
diode-pumped, solid-state, Nd:YAG laser (Coherent Laser Inc.),
emitting 250 mW of CW optical power at a wavelength of ${\rm 1.3\ \mu
m}$. However, in most parts of the experiment, light intensity was
attenuated by a factor of 3 using an air-space-gap fiber-optic coupler
to avoid the saturation of the photoreceivers and to reduce the
possible thermal loading of the EO setup. Active noise suppression
electronics was incorporated in this laser to remove the relaxation
noise. Beyond 5 MHz rf frequency, the laser noise was ${\rm \sim1\
dB}$ above the shot-noise. The polarization purity of the light source
had an extinction ratio of ${\rm >10^4}$ at the output end of the
polarization-maintaining (PM) fiber. The light was then coupled to a
vacuum sealed PM fiber collimator where the output polarization was
rotated to ${\rm +45^o}$ to the azimuthal. The polarization purity
dropped to ${\rm \sim10^2}$ after one fiber coupling. A ${\rm
90^o}$-keyed fiber-optic coupler was used to rotate the input
polarization to the EO crystal from ${\rm +45^o}$ to ${\rm -45^o}$ as
indicated in Fig. 2(b). The collimated ${\rm 0.4\ mm}$ diameter light
beam was sent to the bottom half of the ${\rm LiNbO_3}$ EO crystal
mounted on a ceramic holder that has a clearance hole of 6.35 mm for
the electron beam, see Fig. 1(b). The size of the EO crystal was 6.5 x
2.2 x 1 mm; the optical z-axis (extraordinary axis) was aligned
azimuthally and the x-axis (ordinary axis) was parallel to the
propagation direction of the electron beam. Fluorescent material was
placed around the ${\rm 45^o}$ facet of the ceramic for guiding the
electron beam through the EO crystal. A CCD camera viewed the
fluorescence due to the electron beam from directly above the setup. A
${\rm 45^o}$ pop-up flag with the same fluorescent material was
located 23 cm downstream of the crystal for precise electron beam
location and profile measurements. Each electron beam profile was
recorded and overlaid in Fig. 2(a). Three beam profiles where the
electron beam traversed the EO crystal did not show up clearly on the
flag and their beam position was estimated from the position
dependence of the dipole pitching magnet current. A representative
beam profile shows the electron beam cleared the top portion of the EO
crystal. 

To linearize the modulation and balance the residual birefringence of
the EO crystal, the ${\rm \lambda \over 4}$ waveplate was adjusted so
that the EO system was optically biased at the quadrature
point. Therefore, the resulting electric field-induced optical
modulation constantly rode on a large dc light level. However, only
the transient component of the optical signal was detected by the
optical receiver with the corresponding dc level kept below
saturation. An analyzer crossed at ${\rm -45^o}$ to the input
polarization was positioned after the crystal. A vacuum sealed
multimode (MM) fiber collimator collected the intensity light output
after the analyzer. The light throughput of the complete EO setup was
${\rm\sim12\%}$, with typically 5 mW of optical power received by the
photoreceiver. The laser source was placed inside the concrete
surrounded experimental hall near the EO setup to maintain the high
quality of the PM light, but the output light was transmitted by a
40-meter long MM fiber to the optical receiver outside the
experimental hall for detection and analysis. Light intensity output
from the photoreceiver was sent to a digitizing oscilloscope, each
signal trace was accumulated in 16 to 64 signal averages.  During the
course of the experiment, oscilloscopes with bandwidths of 1-GHz,
3-GHz, and 7-GHz were used in combination of either a 1-GHz of 12-GHz
photoreceivers.  

The electron beam source is the 45 MeV electron beam at the Brookhaven
National Laboratory Accelerator Test Facility (ATF). A drawing of the
ATF layout and its beam lines is shown in Fig. 3.  
A 5 MeV electron beam from a rf photocathode gun was injected into a
linac to boost its energy to 45 MeV. The final beam contained up to
0.6 nC charge in a focused beam diameter of ${\rm\sim0.5\ mm}$ in 10
ps duration at a repetition rate of 1.5 Hz. It was scanned vertically
over a range of a few mm from the bottom of the EO crystal to the top
of the opening by adjusting the driving current of a dipole pitching
magnet. A stable trigger signal synchronized to the electron beam was
obtained from a stripline detector upstream, also depicted in Fig. 3.  

\section{Results}
 
The electron beam induced EO signal was confirmed by a few control
experiments. (1) No photons with wavelengths other than the input
laser were received by the photoreceiver. Such photons may originate
from  nonlinear optical processes as well as transition or Cerenkov
radiation.  
(2) The signal vanished in the absence of the electron beam or the laser beam. 
(3) The signal polarity changed sign when the direction of the induced
electrical field was reversed, or  
(4) when the input laser polarization was rotated by ${\rm 90^o}$. 
The results of (3) are shown in Fig. 4(a) where the electron beam was
steered above or below the EO crystal inducing opposite electric
fields at the path of the laser light causing reversal of the signal
polarity. Figure 4(b) also shows similar polarity flip when the input
polarization was changed from ${\rm +45^o}$ to ${\rm -45^o}$ by using
a ${\rm 90^0}$-keyed fiber-optic coupler. We note that the
polarization of the input light is maintained when it is coupled
either to the fast-axis or the slow-axis of a PM fiber. The insets of
Figs. 4(a), and 4(b), where I(t) in Eq.(\ref{EQN(I(t))}) is plotted,
illustrate the simple intuitive explanation of these changes in EO
signal polarities. When one operates the EO device on the positive
slope of its response function, the polarity of the modulated signal
follows the input. However, when the operation moves to the negative
slope of its response function, that is equivalent to switching the
input polarization from ${\rm +45^o}$ to ${\rm -45^o}$, as shown in
Fig. 2(b), the polarity of the modulated signal becomes opposite to
the input. It is important to point out that all signal traces with
negative (positive) polarity correspond to light intensity drop
(increase). The polarity reversal gives conclusive evidence of the
signal being electro-optical in origin. 

The EO signal dependence on electron beam charge was also
investigated. The electron beam charge was varied by adjusting the UV
intensity irradiating the photocathode of the 5 MeV rf electron
gun. The actual charge was measured by a Faraday cup before the linac
and also by a stripline detector after the linac. Each horizontal
error bar displayed in the inset is the difference between these two
measuring devices, while each vertical error bar is the standard
deviation of 6 sets of signal traces for each charge. The electron
beam position was locked at -1.17 mm away from the laser beam path and
it clearly passed below the EO crystal unobstructed. Individual signal
traces for 5 different charge are shown in Fig. 5, and a linear
$\chi$-square fit to the signal strength is shown in the inset. A
linear dependence of the EO signal with charge was established.  

EO signal dependence on electron beam position was also
investigated. Figure 6(a) displays five signal traces when the
electron beam was steered vertically toward but not traversing the EO
crystal. Each electron beam position was indicated in Fig. 2(a) and
their amplitudes is plotted against their corresponding distance from
the center of the optical beam path in the inset. A $\chi$-square fit
of the data favors equally a $ 1 \over \sqrt{r} $ or a $ 1 \over
{r+a}$ dependence, where $a$ is a constant equal to 1.75 mm. On the
contrary, the same $\chi$-square fit gives a much lower confidence
level on a $ 1 \over r^2 $ or a linear dependence. Therefore, we can
conclude that the EO signal behaves very close to but not exactly as
predicted in Eq.(\ref{final-EO}). The discrepancy needs further
investigation.  

When the electron beam was close to the EO crystal, at beam position
-0.64 mm, a distinctive positive signal with a long ${\rm\sim100\ ns}$
decay time superimposed on the negative EO modulation was
observed. This observation is an indication of the electron beam
partially impinging on the EO crystal. A partially blocked electron
beam profile observed in Fig. 2(a) also supports this argument. To
examine this further, the electron beam was steered to impinge on the
EO crystal and traverse the optical beam path completely. Figure 6(b)
shows these signal traces where the time and the amplitude of the
signal have both been expanded. As the electron beam approached the
optical beam path traversing the EO crystal, the strength of the
positive signal increased and then became negative after passing the
optical beam path. It is conceivable that the electron beam ionizes
the ${\rm LiNbO_3}$ creating electron-hole pairs. Since the mobility
of ions is small compared to the electrons, a transient ion field
remains which produces a EO signal opposite to that due to the
electron beam field. It's decay time will be dictated by the
electron-hole recombination time of the EO crystal.\cite{Taylor}
Therefore, when the origin of the ion field is moved from below to
above the laser beam path, that is from beam position -0.17 mm to
+0.08 mm, the EO signal due to the ionization also changes
polarity. Consequently it provides an unique method to locate the
exact electron beam position with respect to the laser beam
position. However, this ion field fails to diminish when the electron
beam continues to move toward the top of the EO crystal as indicated
by the data trace obtained at the 0.33 mm beam position in
Fig. 6(b). In the present EO design, signal with negative polarity
also corresponds to an intensity drop. Therefore, when the electron
beam strikes the optical beam path, substantial temporal opacity may
be created enhancing the actual strength of the ion
field. Nonetheless, the ion field disappears and the electron beam
field prevails when the electron beam clears the top of the EO
crystal, as shown in Fig. 4(a). 

Since the optical modulation is of electro-optical origin which has a
response faster than the electron pulse duration, the measured
temporal duration is then limited mostly by the bandwidth of the
measurement system and the modal dispersion of the 40-m long out-going
MM fiber. The latter effect was independently measured to have
negligible temporal broadening on the time scale of interest in this
experiment. Figure 7 shows the shortest signal pulse width of 70 ps
recorded on a 7-GHz oscilloscope using a 12-GHz optical receiver. The
instrument response of the same measurement system is displayed in
dashed line. It is worth pointing out that the instrument response was
obtained with a mode-locked IR laser pulse of ${\rm\sim15\ ps}$
duration. The risetime and the pulse width of both the instrument
response and the EO signal traces are comparable, suggesting that the
electron bunch can be inferred to be on the order of ${\rm\sim 15\
ps}$.  

\section{Conclusions}
 
The effectiveness of a Pockels cell field sensor has been demonstrated
for noninvasive measurement on the bunch length of an ultrashort
relativistic electron beam. The signal strength is shown to increase
linearly with the charge and decrease inversely with the distance
between the laser and the electron beam. Currently the temporal
resolution is limited primarily by the detection technique. Although
these results are encouraging, at present the EO modulation is at best
a few percent of the unmodulated dc light level. Methods to improve
the strength of the EO signal and the signal-to-noise ratio are needed
to make it more practical. Nevertheless, EO sensor is clearly an
attractive candidate to explore the ultrashort particle bunch duration
down to the sub-picosecond regime. Measurement of the EO signal using
a 2 ps (or a 0.5 ps) resolution limited streak camera is currently
underway. Using an upgraded pump-probe EO detection scheme and
state-of-the-art ultrafast optical pulse measurement techniques such
as frequency-resolved optical gating or spectral phase interferometry
for direct electric-field reconstruction, relativistic femtosecond
electron bunch may be measured. Furthermore, one can in principle
construct a 2-dimensional ultrafast detector array based on the EO
technique to measure the location, spatial, and temporal profile of
the charged particle beam. Because the EO modulated signal polarity
depends on the induced field direction, the technique is effective for
both positive and negative charged particles, that is electrons as
well as protons and ions.  

\section{acknowledgments}
We wish to acknowledge the support and encouragement of 
Xiejie Wang, Ilan 
Ben-Zvi, Vitaly Yakimenko, Howard Gordon, Mike
Murtagh and Tom Kirk. The efforts
of Victor Usack were essential to our progress.

This manuscript has been written under contract DE-AC02-98CH10886 with 
the U.S. Department of Energy.

\bigskip
\newpage

\section{Figure Captions}

\bigskip

\begin{figure}[ht]
\caption{Experimental setup, (a) showing PM optical fiber input on the
right hand side, followed by the EO crystal and its holder,  ${\rm
\lambda}$/4 waveplate, analyzer position at ${\rm 45^o}$ crossed to
the input polarization, and finally the signal collection multimode
fiber, all mounted on an aluminum base plate anchored to a standard
vacuum flange. (b)  Expanded view of the ceramic holder for the EO
crystal. Fluorescent material is placed at various locations of the
ceramic holder for on-line guiding of electron beam to the EO
crystal.} 
\label{fig1} 
\end{figure}

\begin{figure}[ht]
\caption{(a) Schematic drawing of the EO crystal. A 6.35 mm diameter
clearance hole on the ceramic holder is also shown. Electron beam
propagates along the x direction into the paper.  Several
representative electron beam profiles and their locations with respect
to the laser beam position at z=0 are overlaid to show the maneuver of
the electron beam along the z-axis passing above, through, and below
the EO crystal. Three electron beam positions that were blocked by the
EO crystal did not show up clearly on the flag but is illustrated in
the figure by their beam positions relative to the laser beam
path. Their approximate positions were determined by the pitching
current of the dipole magnet. (b) Schematic cross-sectional view of
the EO crystal (1.0 x 6.5 x 2.2 mm). The electron beam propagates
along the ordinary x-axis, and the transient electric field is induced
along the extraordinary z-axis of the EO crystal. The laser propagates
along the negative y-axis with a collimated beam diameter of 0.4 mm,
its input polarization is oriented either at ${\rm +45^o}$ or ${\rm
-45^o}$ to the extraordinary axis.} 
\label{fig2} 
\end{figure}

\begin{figure}[ht]
\caption{Accelerator Test Facility (ATF) beam lines. Also indicated
are the locations of the EO experiment setup at beam line \#3 and the
trigger signal extracted from a stripline detector at the linac
section. Electron beam travels from right to left.} 
\label{fig3} 
\end{figure}

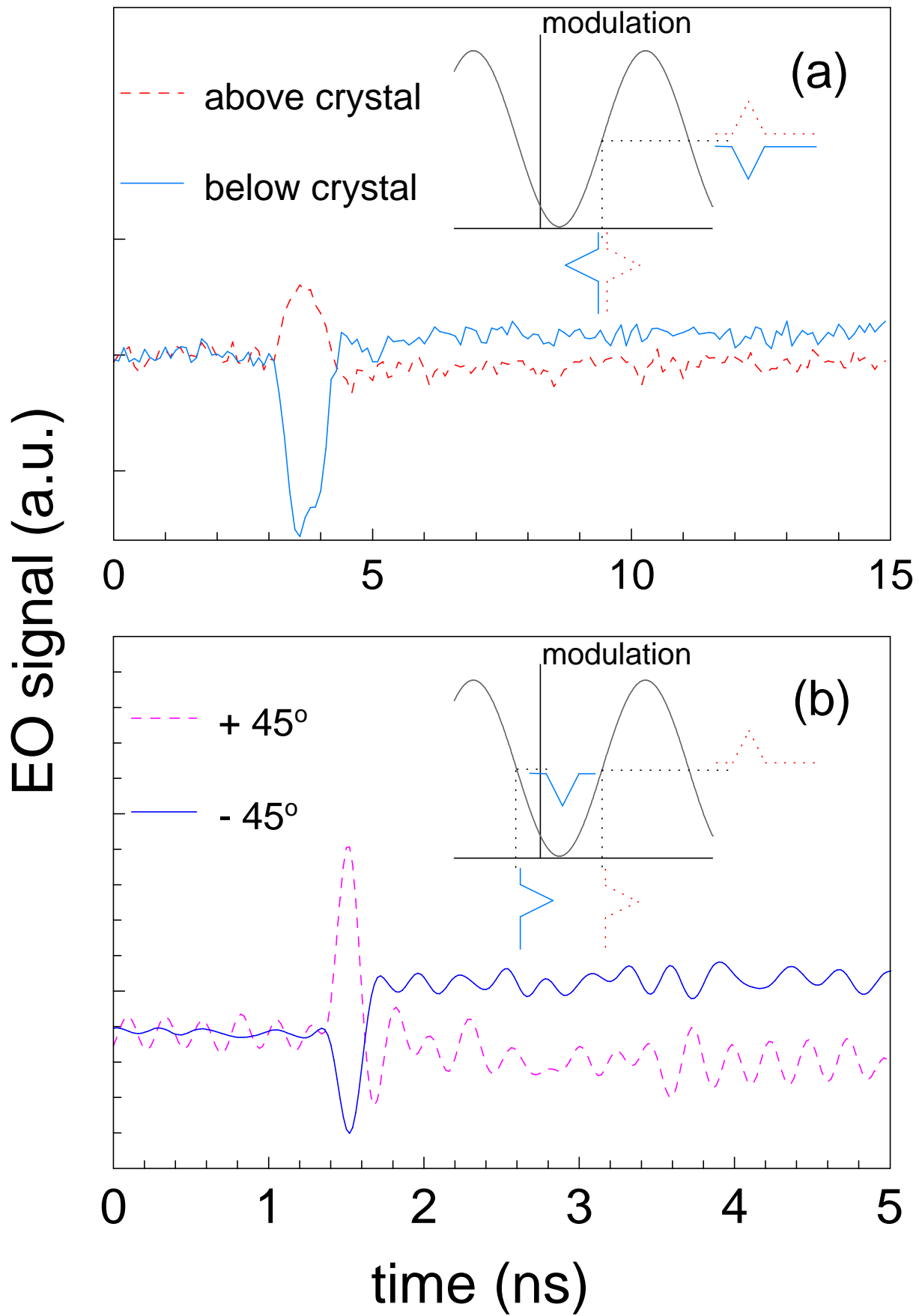
\begin{figure}[ht]
\caption{(a) EO modulated signal when the electron beam passed
unobstructed below (solid line) and above (dashed line) the EO crystal
so that the induced electron beam field is reversed.   
(b) EO modulated signal when the input laser polarization was flipped
from ${\rm+45^o}$ to ${\rm -45^o}$. Note that the time scale of the
two plots are different because a 1-GHz and a 12-GHz   
photoreceiver was used in (a) and (b) respectively. 
A pictorial representation of the optical launching condition is
illustrated in the in-set of each figure.} 
\label{fig4} 
\end{figure}

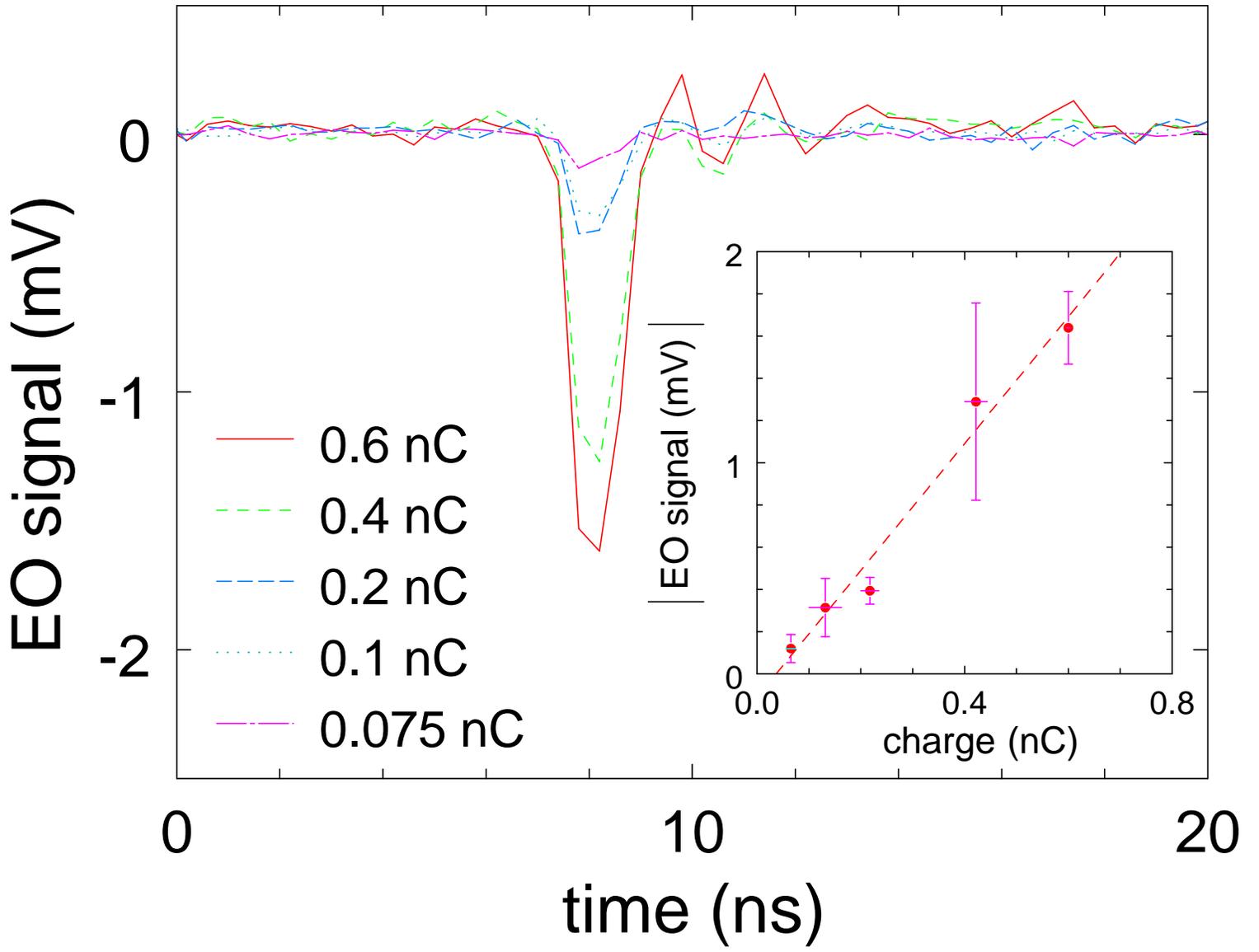
\begin{figure}[ht]
\caption{Increase of the EO modulated signal with electron beam charge.
The electron beam position is locked at -1.17 mm away from the laser
beam path. The inset shows the  dependence of the EO signal with
charge, dashed line is the linear fit to the data.} 
\label{fig5} 
\end{figure}

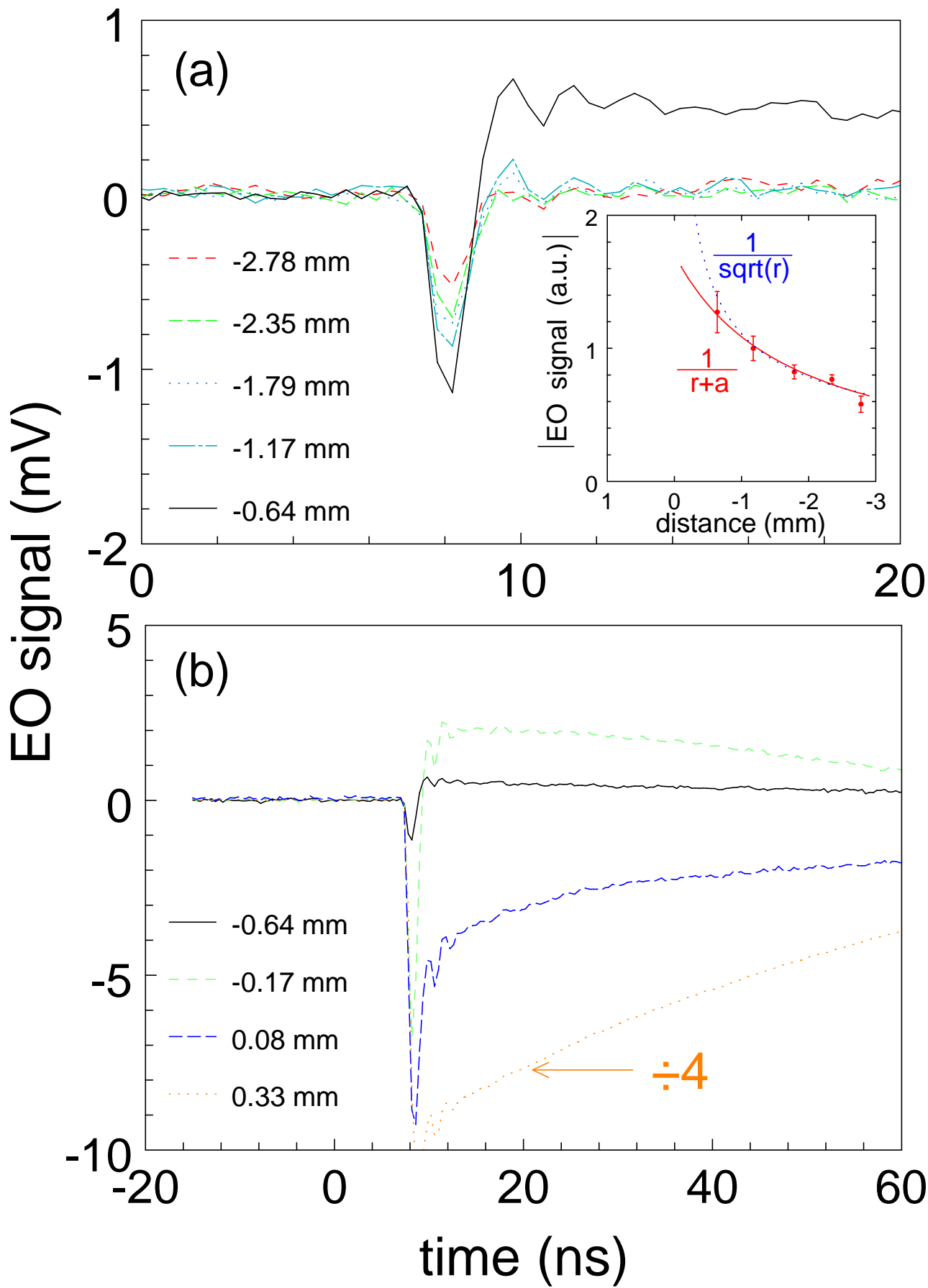
\begin{figure}[ht]
\caption{(a) Increase of the EO modulated signal with electron beam
approaching the EO crystal from the negative z-axis. The inset shows
the signal plotted against the distance of the electron beam away from
the laser beam path.  A $1 \over {r+a}$ and a $1 \over
\sqrt{r}$dependence is also fitted into the data.  
(b) Same as (a) but when the electron beam was moved to irradiate on
the EO crystal and traverse the laser beam path.  
See Figure 2 for detailed electron beam positions. Note that the
signal trace of -0.64 mm is plotted on both figures for comparison and
the data trace of 0.33 mm was divided by a factor of 4 to fit in the
current vertical scale.} 
\label{fig6} 
\end{figure}

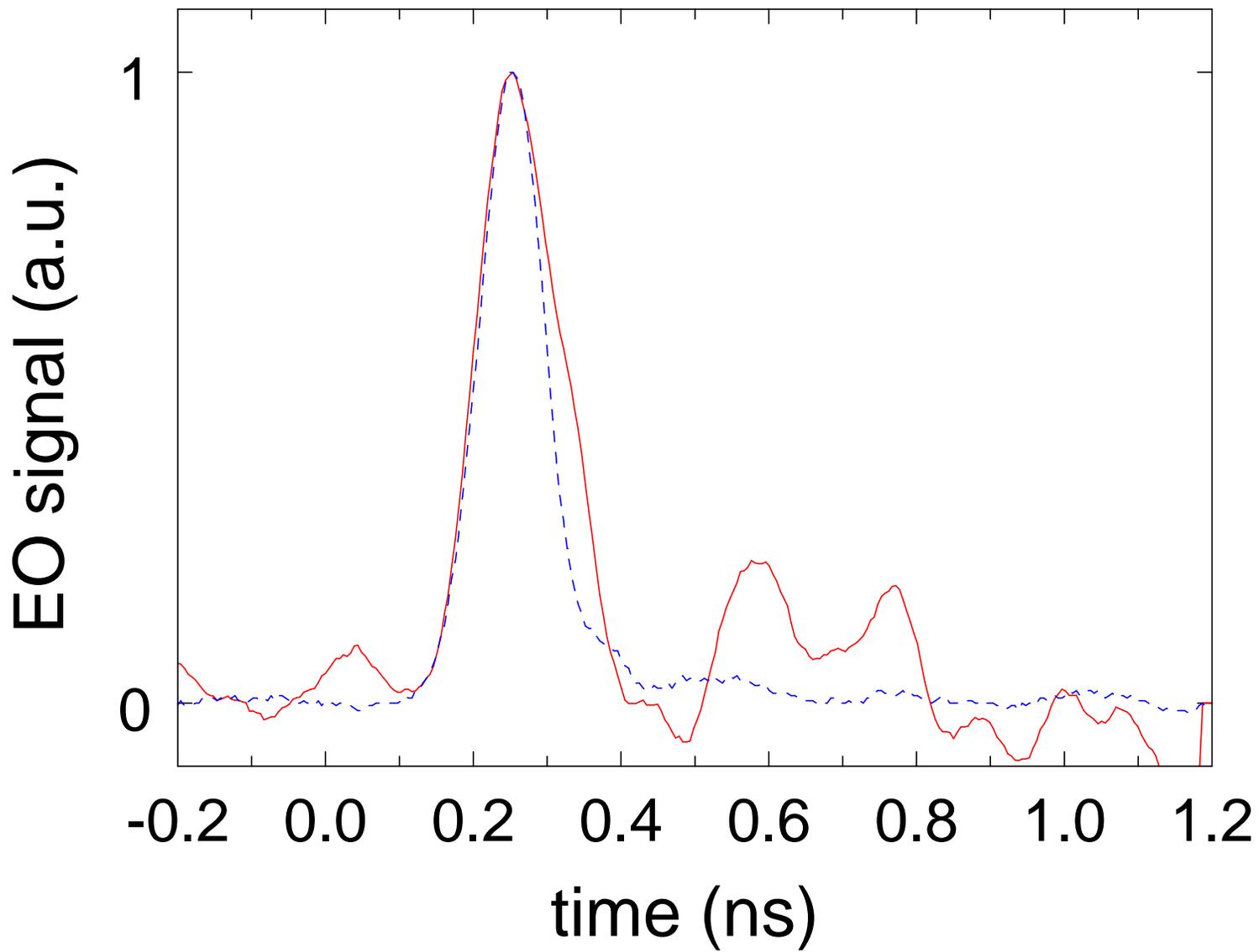
\begin{figure}[ht]
\caption{Solid line - the EO signal detected by a 12 GHz photoreceiver
on a 7 GHz digital oscilloscope. Dashed line - instrument response of
the measurement system using a ${\rm \sim15\ ps}$ IR pulse.} 
\label{fig7} 
\end{figure}

%
%

%
%

\end{document}